\documentclass[lettersize,journal]{IEEEtran}
\usepackage{amsmath,amsfonts}
\usepackage{algorithmic}
\usepackage{algorithm}
\usepackage{array}
\usepackage{subfig}
\usepackage{textcomp}
\usepackage{stfloats}
\usepackage{url}
\usepackage{verbatim}
\usepackage{graphicx}
\usepackage{cite}
\usepackage[acronym]{glossaries}
\usepackage{svg}
\usepackage{float}
\usepackage{booktabs}
\usepackage{makecell}
\usepackage{bm}
\usepackage{balance}

\usepackage{xcolor}
\definecolor{rncolor}{HTML}{B22222} 

\usepackage{color}

\usepackage{xspace}
\newcommand{\eg}{\textit{e.g.},\xspace}
\newcommand{\ie}{\textit{i.e.},\xspace}

\usepackage{soul}
\usepackage{tikz}
\usetikzlibrary{arrows.meta, positioning}

\newcommand{\ac}[1]{\gls{#1}}
\newcommand{\acp}[1]{\glspl{#1}}

\usepackage{tikz}
\DeclareRobustCommand{\circledchar}[1]{%
  \tikz[baseline=(char.base)]{
    \node[shape=circle, fill=black, text=white, inner sep=0.5pt] (char) {\fontfamily{phv}\fontsize{7.5}{9}\selectfont\textbf{#1}};
  }%
}
\DeclareRobustCommand{\circlednum}[1]{%
  \tikz[baseline=(char.base)]{
    \node[shape=circle, draw, inner sep=0.5pt] (char) {#1};
  }%
}

\begin{document}

\title{Automated Heuristic Design for Network Operations}

\author{Reza Namvar, \IEEEmembership{Student Member, IEEE}, José Gallego, \IEEEmembership{Student Member, IEEE}, Jose A. Ayala-Romero, \IEEEmembership{Member, IEEE}, Livia Elena Chatzieleftheriou, \IEEEmembership{Member, IEEE}, Andres Garcia-Saavedra, \IEEEmembership{Senior Member, IEEE}, Albert Banchs, \IEEEmembership{Senior Member, IEEE}, and
Marco Fiore, \IEEEmembership{Senior Member, IEEE}
\thanks{R. Namvar, J. Gallego, and A. Banchs are with IMDEA Networks Institute, Madrid, Spain and Universidad Carlos III de Madrid, Spain. E-mail: \{reza.namvar, jose.gallego, albert.banchs\}@networks.imdea.org}
\thanks{J. A. Ayala Romero, and A. Garcia-Saavedra are with NEC Laboratories Europe, Heidelberg, Germany. E-mail: \{andres.garcia.saavedra, jose.ayala\}@neclab.eu}
\thanks{L. E. Chatzieleftheriou is with the Technical University of Delft, Delft, the Netherlands. E-mail: L.E.Chatzieleftheriou@tudelft.nl}
\thanks{M. Fiore is with IMDEA Networks Institute, Madrid, Spain. E-mail: marco.fiore@networks.imdea.org}
}




\maketitle

\newacronym{ai}{AI}{Artificial Intelligence}
\newacronym{ahd}{AHD}{Automated Heuristic Design}
\newacronym{ldpc}{LDPC}{Low-Density Parity Check}
\newacronym{ml}{ML}{Machine Learning}
\newacronym{cnu}{CNU}{Check Node Update}
\newacronym{vnu}{VNU}{Variable Node Update}
\newacronym{tb}{TB}{Transport Block}
\newacronym{llr}{LLR}{Log Likelihood Ratio}
\newacronym{ber}{BER}{Bit Error Rate}
\newacronym{gp}{GP}{Genetic Programming}
\newacronym[plural=LLMs,firstplural=Large Language Models (LLMs)]{llm}{LLM}{Large Language Model}
\newacronym{eoh}{EoH}{Evolution of Heuristics}
\newacronym{revo}{ReEvo}{Reflective Evolution}
\newacronym{bp}{BP}{Belief Propagation}
\newacronym{snr}{SNR}{Signal-to-Noise Ratio}

\begin{abstract}
Network operation relies on heuristics to solve many tasks rapidly and efficiently across the protocol stack. These heuristics are the result of thorough human-driven design rooted in expert knowledge of the target system and problem. 
Recently, approaches powered by Artificial Intelligence have shown promising results in devising solutions that outperform long-established heuristics in classical problems.
We explore the possibility of applying such Automated Heuristic Design (AHD) frameworks to network environments by ($\bm{i}$) discussing the general integration of AHD with network operation and the associated challenges, as well as ($\bm{ii}$) proposing a practical implementation of AHD for a specific networking task, \ie 
5G decoding. Initial results show how modern AHD tools can devise heuristics for Low-Density Parity Check decoding 
on par with state-of-the-art solutions implemented in production systems.
\end{abstract}



\section{Introduction}

Heuristic algorithms, \ie fast, solutions that lack formal performance guarantees but effectively solve NP-hard optimization tasks, are the cornerstone of industry-grade network operation and provide practical solutions for signal detection, channel estimation, scheduling, resource allocation, routing, or congestion control.
Traditionally, heuristics are the result of thoughtful manual design driven by deep domain knowledge, leading to a combination of computational efficiency and engineering accountability that are hard to beat. In fact, \ac{ai} models proposed during the past decade to address the same tasks, \eg~\cite{learning_heu,computers_can,towards_provably,constrained,robust,physics}, have largely failed to supplant long-established heuristics in production networks due to higher complexity and lack of transparency.

\textbf{Automated Heuristic Design.}
\ac{ahd} is a different paradigm through which \ac{ai} can move beyond conventional human-designed heuristics for networked systems. Unlike legacy applications of \ac{ai} that employ complex computational models trained on massive data to solve the target task, \ac{ahd} leverages \ac{ai} to \textit{design a simple heuristic} for the task---thus retaining the efficiency and accountability  properties inherent to heuristic algorithms,  while benefiting from the unparalleled proficiency of \ac{ai} in devising solutions with higher performance than human-defined programs.

Originally based on \ac{gp},~\ac{ahd} has been  revolutionized by the advent of \acp{llm}, which have replaced the rigid search spaces and limited expressiveness of \ac{gp} with flexible, natural language-driven code generation.
Specifically, recent frameworks have successfully integrated \acp{llm} into evolutionary loops that allow refining candidate heuristics while avoiding hallucination or suboptimal code generation.
Seminal frameworks such as FunSearch~\cite{funsearch} demonstrated that pairing \acp{llm} with programmatic evaluators in distributed, island-based architectures and best-shot prompting
can lead to novel mathematical knowledge.
Subsequent proposals like \ac{eoh}~\cite{eoh} and \ac{revo}~\cite{reevo} have enhanced \ac{ahd} through ($i$)~tailored prompting strategies that control the trade-off between exploration and exploitation, ($ii$)~dual-evolution that co-evolves the algorithmic code and its corresponding natural language explanation, or ($iii$)~verbal gradients that emulate human expert reflection to provide natural-language feedback across code generations.
%

Modern \ac{ahd} frameworks based on \acp{llm} prove capable of devising new heuristics that are human-interpretable and attain significantly better performance than state-of-the-art solutions designed by domain experts for classical problems like cap set~\cite{funsearch}, bin packing~\cite{funsearch}, traveling salesman~\cite{eoh}, flow-shop scheduling~\cite{eoh}, or circle packing~\cite{llm4ad}.

\textbf{\ac{ahd} for networking.}
Motivated by the success of the paradigm in the context of optimization tasks, we discuss in this paper the application of \ac{llm}-based \ac{ahd} to network systems.
In fact, using \ac{ahd} frameworks to solve networking tasks is not straightforward and requires addressing novel shortcomings.
First, \ac{ahd} has been validated almost exclusively on static combinatorial optimization problems where objectives are well defined, evaluations of the solution quality are inexpensive, and thousands of runs per candidate are feasible; this is not the case in network systems where metrics are multi-dimensional and stochastic, and performance needs to be assessed via computationally-heavy simulation or time-consuming experiments.
Second, existing \ac{ahd} tools are often over-specialized to their target problem structures, lacking versatility across the wide range of heterogeneous tasks encountered in networks. 
%
Third, public implementations of state-of-the-art \ac{ahd} frameworks are often incomplete, only providing high-level operation pipelines while lacking detailed file structures and practical specifications required for deployment in complex networking environments.

Ultimately, directly using existing FunSearch-, EoH-, or ReEvo-style pipelines to generate an heuristic for a network task is not possible.
Instead, the overall \ac{ahd} methodology must be revisited, from how heuristics are embedded into the system to how scores are constructed and how evaluations are scheduled under strict resource constraints.



\textbf{Our contributions.}
We make a first effort to overcome the challenges above and adapt \ac{llm}-based \ac{ahd} to networking scenarios. 
Our contributions are summarized as follows.
 
        

    \begin{itemize}
        \item We propose a complete four-phase methodology to tailor \ac{ahd} for tasks emerging in the networking domain. 
        \item We present one practical application case study, adapting an \ac{ahd} engine to solve a specific sub-function of \ac{ldpc} processing, showing results comparable to state-of-the-art human-designed heuristics.
        \item We open source our full implementation of a \ac{ahd} tool for networking, fostering future research on the topic. 
    \end{itemize}




\section{Heuristic Generation for Networks}
\label{sec:heuristic_generation_for_networks}

\begin{figure*}[t]
    \centering
    \includegraphics[width=0.9\linewidth]{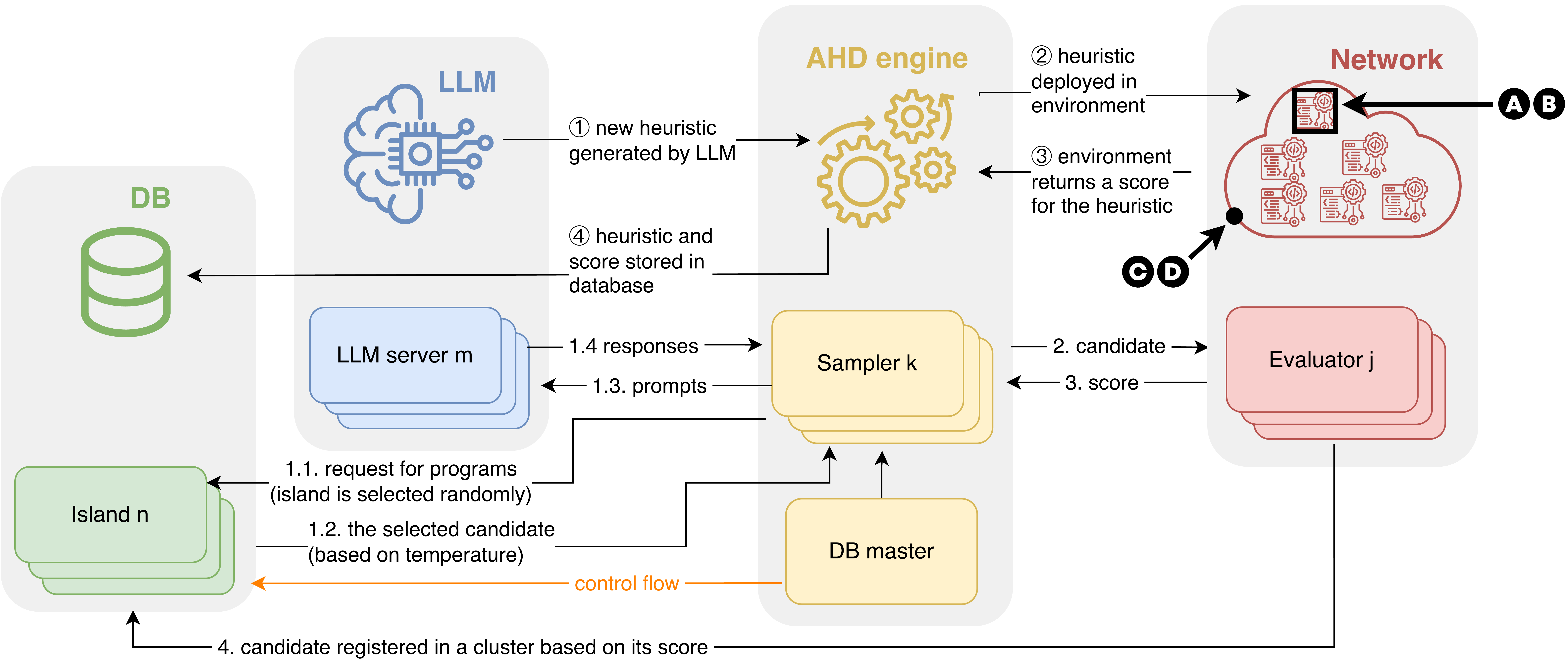}
    \caption{Overview of \ac{ahd} applied to network systems, with high-level logical blocks and their implementations.
    }
    \label{fig:ahd_networking}
\end{figure*}

The proposed operation of \ac{ahd} for network systems is illustrated in Fig.~\ref{fig:ahd_networking}. The \ac{ahd} engine leverages modern \acp{llm} to produce heuristics for a specific networking tasks in step \circlednum{1}. The heuristics are then deployed in the relevant network environment in step \circlednum{2} and tested for compatibility with system requirements and general performance, returning a score to the \ac{ahd} engine in step \circlednum{3}. Best-performing heuristics are stored in a database to track their evolution and retain the final code in step \circlednum{4}.

Implementing the general flow above requires a dedicated methodology to address four specific design phases, which we denote as A--D in Fig.~\ref{fig:ahd_networking} and discuss next.

\subsection{\ac{ahd} Suitability Analysis}
Given a target networking problem, the first phase consists in assessing if the task satisfies the criteria that would make \ac{ahd} an adequate tool for solving it, as follows. 


    
    \textbf{Computational intractability.} The problem must be sufficiently complex that exact methods (\eg polynomial-time optimal solutions to MILP formulations) are infeasible for real-time operation, hence calling for a heuristic approach.
    
    \textbf{Evaluation efficiency.} The \ac{ahd} loop shall evaluate thousands of candidate programs hence the problem must allow for rapid assessment of solution quality. Lightweight, high-throughput simulation or experimental environments must be available that accurately capture the performance metrics.
    
    \textbf{Manageable stochasticity.} While networking environments are inherently stochastic, excessive variance in evaluation metrics can mask the scoring values of a generated heuristic. 
    With high noise floor, the evolutionary process may fail to distinguish between superior logic and lucky random seeds.

\subsection{Atomic Task Selection}
\label{subsec:atomic_task_selection}
Networking functions are often too complex to be addressed as a whole by \ac{ahd}. The second fundamental stage of our approach
is the decomposition of the target network function logic into a fixed \textit{skeleton} and an evolving \textit{atomic task}. The atomic task represents the specific algorithmic logic to be optimized by the \ac{llm}-based \ac{ahd}. To ensure successful evolution, the atomic task must adhere to the following principles.

    \textbf{Algorithmic logic over parametric tuning.} The target must be the discovery of symbolic logic (\ie code structure) and not merely numerical values. If the goal is just optimizing scalar weights, traditional techniques (\eg Bayesian optimization) are more efficient. In contrast, \ac{ahd} is intended for the generation of novel algorithmic logic.
    
    \textbf{Syntactic simplicity, semantic complexity.} The atomic task should be syntactically concise, \ie short enough to fit comfortably within the \ac{llm}'s context window and minimize hallucination risks, while governing complex system behaviors. A common example in the literature is \textit{priority functions}: 
    for instance, in scheduling, a concise mathematical expression as a priority function that just sorts users or packets, captures well the complex macroscopic network performance.
    
    \textbf{Structured state representation.} The input to the atomic task must be structured numerical or categorical data (\eg signal-to-noise ratio, queue depth, latency requirements). Unstructured inputs, such as raw natural language system logs, are generally unsuitable for the high-frequency numerical reasoning required in the context of \ac{ahd}.
    
    \textbf{Functional purity.} The atomic task should ideally be stateless and self-contained. It must act as a deterministic mapping from inputs to outputs without side effects or dependencies on global variables. This isolation ensures that the evaluation is robust and that the \ac{llm} focuses solely on the transformation logic rather than state management.

\subsection{Scoring Function Design}
The design of the scoring function is the paramount third stage of our methodology, providing the ground truth that guides the \ac{ahd} evolutionary process. In networking contexts, complex, multi-objective performance metrics shall be condensed into a scalar value that reflects all system priorities.

    \textbf{Hierarchical objectives and soft constraints.} Networking problems inherently involve multiple hierarchical objectives. For instance, satisfying reliability constraints (\eg outage probability) 
    should take 
    precedence over service maximization (\eg user fairness or throughput). To capture this, the scoring function must effectively enforce \textit{soft constraints}.
    This is done via linear scalarization or non-linear aggregation.
        In \textit{linear scalarization}, a weighted sum approach 
        approximates hierarchical preferences; however, this requires careful tuning of weights across different orders of magnitude to ensure that violations of primary constraints (\eg outage) strictly dominate gains in secondary metrics.
        In \textit{non-linear aggregation}, 
        product-based formulations 
        naturally penalize performance when constraints are violated; here, normalization of input metrics is critical to prevent numerical instability.
    
    \textbf{Dense versus sparse feedback.} The scoring feedback must be continuous (dense) rather than binary (sparse). A discrete function (\eg returning 0 if an outage occurs, 1 otherwise) fails to provide the ``gradient'' necessary for the \ac{llm} to distinguish between a near-miss solution and a catastrophic failure. A continuous penalty function allows the evolutionary search to incrementally converge toward feasible regions.

\subsection{Evaluation Methodology}
Heuristic evaluation presents unique challenges in the networking domain compared to pure mathematical discovery. Unlike static combinatorial problems 
governed solely by dimensionality, network functions are defined by high-dimensional, stochastic state spaces including, \eg topology, traffic patterns, user mobility, or channel propagation.
This introduces a fundamental trade-off between \textit{generalization} and \textit{computational cost}, as follows.

\textbf{Generalization versus simulation time.} Assessing the robustness of a heuristic requires averaging performance over a large distribution of network scenarios and conditions,
increasing the evaluation latency per candidate. Excessive simulation time can bottleneck the evolutionary loop.
    
\textbf{Specialization as a feature.} Conversely, narrowing the evaluation to a specific network configuration risks overfitting, which, can be re-framed as \textit{specialization} in the context of \ac{ahd}. Since generated heuristics are lightweight code snippets, it is feasible to deploy a library of specialized algorithms tailored to diverse network conditions (\eg different, dedicated schedulers for high-mobility and near-static users), rather than relying on a single ``one-size-fits-all'' solution.


\section{\ac{ahd} in Action} 
\label{sec:AHD_for_LDPC_our_solution}



We showcase the proposed \ac{ahd} framework in a representative networking use case, \ie \ac{ldpc} decoding. We first present the target task, then describe our \ac{ahd} implementation of phases A--D, and finally discuss system settings and results.

\subsection{An Introduction to \ac{ldpc} Decoding}

\begin{figure}[t]
    \centering
    \includegraphics[width=1\linewidth]{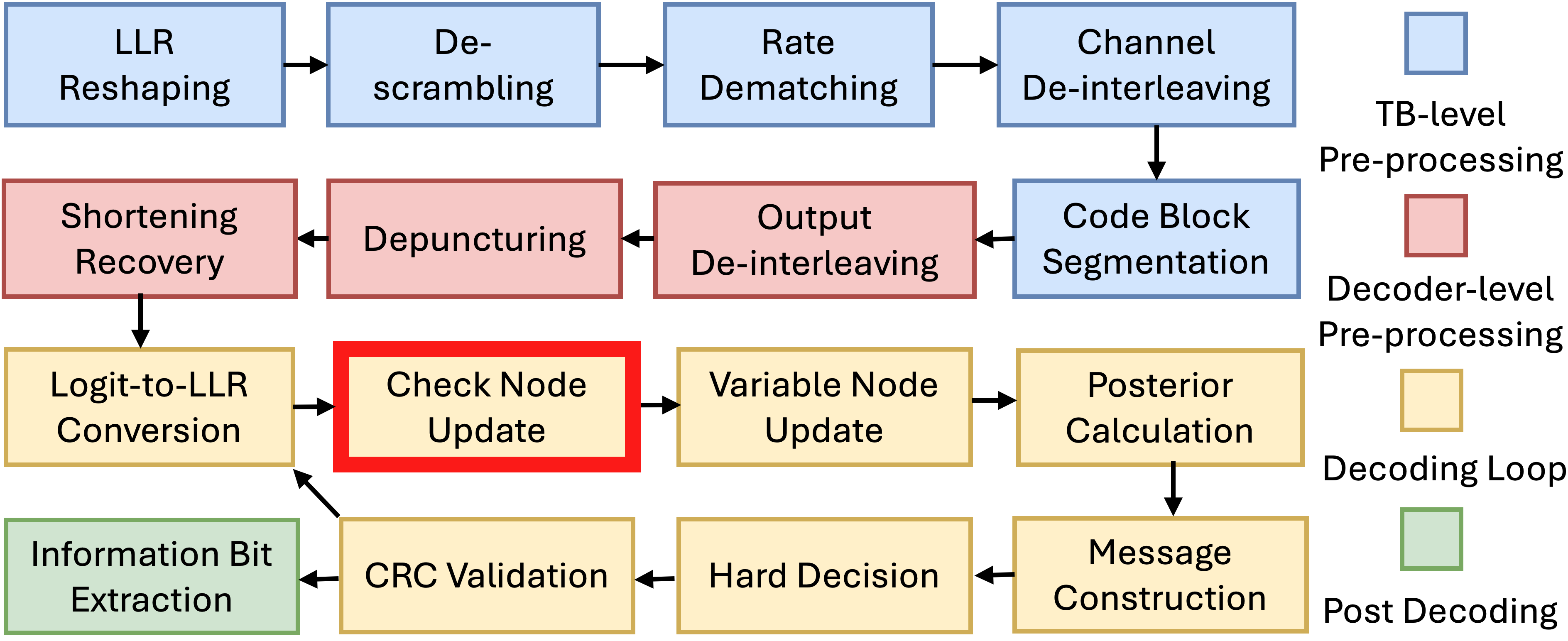}
    \caption{5G LDPC decoding pipeline, highlighting the atomic task considered for \ac{ahd}.}
    \label{fig:ldpc_decoding_pipelince}
    \vspace*{-8pt}
\end{figure}

\ac{ldpc} allows recovering a transmitted message from a received (possibly corrupted) signal. In 5G systems, the receiver decodes 5G NR \ac{tb}s in four conceptual stages, summarized in Fig.~\ref{fig:ldpc_decoding_pipelince}.

\textbf{\ac{tb} pre-processing.} 
A ``soft demapper'' produces a \ac{llr} $L(X)$ per coded bit $X$, \ie a real-valued measure of confidence about bit $X\in\{0,1\}$ given the channel observation $Y$.
Positive $L(X)$ soft values are interpreted as if bit~0 was transmitted, negative values as bit~1, and values close to zero indicate high uncertainty.
Through \emph{\ac{llr} Reshaping}, the incoming soft values are reorganized so that they are grouped per \ac{tb}  and per nominal codeword position, rather than as a flat stream.
\emph{Descrambling} removes the pseudo-random sequence that was applied at the transmitter, so that each \ac{llr} can again be associated with a specific coded bit. 
The \emph{Rate Dematching} operation conceptually reconstructs the circular buffer used by 5G\,NR rate-matching: received bits are placed back into their buffer positions and neutral padding bits with zero \ac{llr} are inserted wherever no observation was transmitted, so that the buffer has the correct length. 
\emph{Channel De-interleaving} then reverses the bit-level interleaver associated with modulation and resource mapping, restoring the ordering in which bits belong to each \ac{tb}. 
Finally, \emph{Code Block Segmentation} slices the reconstructed \ac{tb}s into smaller \ac{ldpc} code blocks, each corresponding to one codeword of a 5G\,NR \ac{ldpc} base graph.

\textbf{5G \ac{ldpc} pre-processing.} 
The above representation is refined at the level of code blocks (\ie grouping multiple bits together),
matching an internal \ac{ldpc} base code. 
An \emph{Output De-interleaving} step reverses the specific 5G rate-matching interleaver, when configured, arranging the bits in the exact order required by the \ac{ldpc} parity-check matrix. 
\emph{Depuncturing} reconstructs punctured parity bits of the \ac{tb}s in the standardized 5G\,NR scheme,
either because their structure is known (header) or because they do not carry additional information (tail). 
%
\emph{Shortening Recovery} handles the “filler bits’’ that were zero-padded before encoding: these bits are reintroduced with LLRs of very large magnitude, representing almost infinite confidence that they are equal to the known filler value. Each \ac{ldpc} code block now appears to the decoder as a complete codeword consistent with the 5G\,NR base graph.

%

\textbf{Core \ac{ldpc} decoding loop}.
This is an iterative \ac{bp} loop 
%
using a Tanner graph, \ie a bipartite graph with ``variable nodes'' (representing message bits) connected to ``check nodes'' (representing parity constraints), and edges connecting each bit to the checks it participates in.
A solution is valid if every bit satisfies its connected parity check, \ie
the bits involved in a parity check sum to an even number. 
Before decoding starts, \emph{\ac{llr} Clipping} constrains the input values to a specific range to prevent numerical errors, and a
\emph{Logit-to-\ac{llr} Conversion} ensures that the signs of the internal definition of $L(X)$ are consistent with the BP update rules. 


Each iteration of the core loop then consists of two main message-passing operations, 
namely \emph{\ac{cnu}} and \emph{\ac{vnu}}, followed by \ac{tb}-level checks. 
In \ac{cnu}, every check node combines the incoming variable-to-check messages from all its neighboring bits and produces outgoing messages back to each of them, 
intuitively computing how well the parity equation associated with that check node is satisfied and pushing each bit towards those values that make the parity more likely to hold. 
Different mathematical rules can be used. 
We elaborate more on this in the next subsection, as this is the core of our \ac{ahd} use case.
%
In the subsequent \ac{vnu}, each variable node collects its incoming check-to-variable messages and combines them with the original channel \ac{llr} for that bit. This results in updated belief about the bit, expressed as an \emph{a-posteriori \ac{llr}}, and new outgoing messages to each neighboring check that exclude the contribution of that specific check (extrinsic information). 
Combined, these operations are summarized as \emph{Message Construction}: channel LLRs and extrinsic information are added to form the messages that will be used in the next BP iteration.

\textbf{CRC-based early stopping}. A \emph{Hard Decision} step maps the a-posteriori \ac{llr} soft values into binary ``hard'' decisions, by comparing each \ac{llr} to zero. 
From these decoded codewords, 
\emph{CRC Validation} is performed: code-block and transport-block CRCs are recomputed and their signatures are checked. 
If CRC Validation succeeds for a given \ac{tb}, that \ac{tb} is marked as decoded and removed from further iterations. 
If, after an iteration, all TBs in the batch have passed CRC Validation, the outer loop stops; 
otherwise, the loop continues until either all TBs are decoded or the maximum iteration budget is reached. 

\textbf{Post-processing.} The final information bits are extracted from the output of the steps above by removing parity bits and any filler or shortened bits from the decoded TBs.

\subsection{Practical \ac{ahd} for \ac{ldpc} Decoding} 



Our \ac{ahd} implementation follows the four phases presented in Sec.~\ref{sec:heuristic_generation_for_networks},
resulting in the distributed and asynchronous \ac{ahd} engine architecture depicted in the bottom part of Fig.~\ref{fig:ahd_networking}.

At its core, our \ac{ahd} tool hinges on the Funsearch model~\cite{funsearch} and is coordinated via shell scripts that automate the startup, monitoring, and synchronization of all distributed components.
The storage for candidate heuristics is split across several islands, each maintaining
its own collection of code snippets and their respective scores, organized into \emph{clusters}. Each cluster stores a distinct set of programs that share the same score. Sampler processes constantly communicate with these islands over HTTP; they request a configurable number of heuristics (sampled from clusters via a temperature-controlled scheme that can bias selection toward higher-scoring clusters as the island grows), augment them with natural-language instructions and other configurable prompt text, and send them to an \ac{llm} running on one (or more) \ac{llm}-server instances. The model returns a modified version of the same function, which the sampler forwards to an evaluator.

Evaluators execute candidates in a sandboxed environment, test them over a fixed set of problem instances, and return a score. If a candidate produces feasible solutions
the corresponding island stores it (placing it into the cluster matching its score) and updates its local population; 
if not 
(\eg due to runtime errors, timeouts, or unstable behavior), it is discarded. In addition to this continuous evolutionary loop, the database master can periodically trigger a lightweight genetic reset step across islands, re-seeding weaker islands from stronger ones according to a configurable reset policy.

All of these components run as separate services, communicate via simple HTTP APIs, and can be deployed either as local processes or as Docker containers. We use Qwen3 as the backbone \ac{llm} for code generation and refinement. 


\textbf{\circledchar{A} \ac{ahd} suitability analysis.}
\ac{ldpc} decoding is one of the fundamental, yet most \textit{computationally complex}, networking problems. 
Moreover, it allows for \textit{evaluation efficiency}, as the scoring function (later detailed) allows so.  
Finally, \textit{manageable stochasticity} is achieved by considering a small granularity in the form that context is taken as input.   

\textbf{\circledchar{B} Atomic task selection.}
We aim for \textit{algorithmic logic over parametric tuning}. 
Targeting the whole \ac{ldpc} processing may result in larger improvements, but risks to be a too large task for \ac{ahd} and curbs linking performance improvements to individual components.
Thus, we focus on a concrete atomic task, \ie the \ac{cnu} function, highlighted in Fig.~\ref{fig:ldpc_decoding_pipelince}, which is the most computationally intensive part of the decoding pipeline.
Standard (hand-crafted) functions include Boxplus (exact sum–product), numerically stabilized Boxplus–$\phi$, or approximations like Min-Sum and Offset Min-Sum. 
By exploring alternative implementations beyond these classical baselines, our goal is to discover new \ac{cnu} functions that offer a superior trade-off between reliability and complexity. 

\begin{figure} 
    \centering
    \includegraphics[width=0.96\linewidth,trim={0 10pt 0 0},clip]{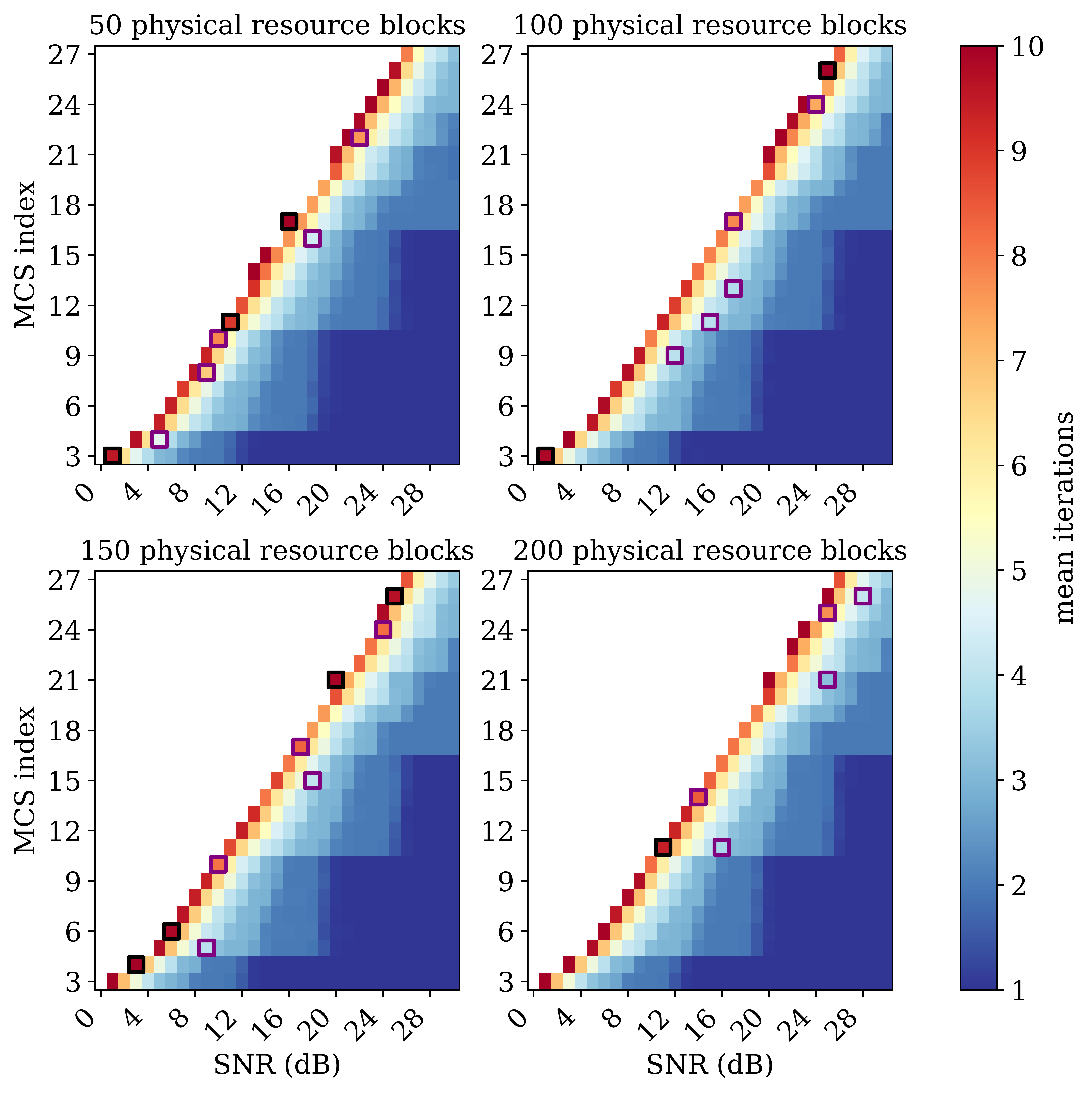}
    \caption{Average number of iterations needed to decode \ac{tb}s across network contexts. White indicates that messages cannot be decoded. The contexts used by our \ac{ahd} are highlighted, in black if not all the \acp{tb} can be decoded, in purple otherwise.}
    \label{fig:heatmap}
    \vspace*{-8pt}
\end{figure}

The \ac{cnu} task has a \textit{structured state representation} and \textit{functional purity}:  
the \ac{ldpc} decoder can operate under many combinations of number of physical resource blocks, Modulation and Coding Scheme (MCS) index, and Signal to Noise Ratio (SNR) measured in dB.
The performance of the decoding is highly dependent on such state, as shown in Fig.~\ref{fig:heatmap} : more complex such ``context'' entail lower success rates in correctly decoding messages and more iterations to do so.

\textbf{\circledchar{C} Scoring function design.}
Our platform 
aggregates the outcome of the CNU simulation into a single scalar score that guides the search. 
The score yields \textit{hierarchical objectives and soft constraints}, as reflects the three main concerns of a practical decoder: reliability, error-rate quality, and complexity.

Reliability is captured by the number of \ac{tb}s that remain undecoded after the maximum number of iterations; heuristics that systematically fail to decode are heavily penalized and quickly discarded. 
Among functions that are able to decode almost all blocks, we then consider the resulting \ac{ber} at the end of the decoding process, so that small gains in reliability obtained at the cost of much worse bit-level decisions are not considered attractive. Finally, for heuristics with comparable reliability and \ac{ber}, we prefer those that reach successful decoding in fewer iterations, which corresponds to lower complexity and latency. The score is constructed so that these three criteria are effectively ordered by priority: avoiding pathological behaviour comes first, then maximizing the number of correctly decoded blocks, then improving \ac{ber}, and only then reducing the iteration count.

In practice, this means that the platform behaves as if it was optimising reliability subject to a minimum quality level, and only afterwards trading off remaining gains against complexity.
We achieve hierarchical objectives and soft constraints by weighting with different multipliers the above factors in the score function: 
$10^9$ for catastrophic failure,
$10^7$ for undecoded \ac{tb}s count,
$10^6$ for the total \ac{ber}, and
$10^0$ for iteration count.
finally, despite the context given as input being naturally discrete (as apparent in Fig.~\ref{fig:heatmap}), our score function considers multiple and dense values where relevant (\eg for the SNR) hence allowing for \textit{dense versus sparse feedback}.

\textbf{\circledchar{D} Evaluation methodology.}
The score function must be computed on a choice of target contexts. Fig.~\ref{fig:heatmap} reveals three operating zones:
($i$)~a straightforward region with low iterations and almost no failures towards the bottom right of each plot; ($ii$)~a hard region with regular decoding failures mapping to the white region in each plot; and ($iii$)~an intermediate band where decoding is often successful but requires many iterations and occasionally fails.
We empirically verify that targeting easy contexts makes the problem undemanding and brings little incentive for \ac{ahd} to discover better heuristics, whereas hard contexts are not informative enough for \ac{ahd} to understand the quality of the solutions it generates---as they tend to yield failures only. Targeting the intermediate band of contexts with mixed decoding results proves the right choice.

%
%

To address the \textit{generalization versus simulation time} trade-off, we test different set sizes of contexts (from 30 to 1, marked by squares in Fig.~\ref{fig:heatmap}) within the scoring function.
Ultimately, we find that using a single context lying near the boundary of the region where decoding is possible is sufficient for \ac{ahd}: indeed, one context in that area makes the task both challenging and informative, creating space for exploration of heuristics that ($i$)~decode more blocks successfully, ($ii$)~reduce the required iterations, or ($iii$)~improve both metrics at once; at the same time, extremely poor rules are immediately identified as they fail on most blocks of the target context.
Selecting a single context also dramatically reduces the memory footprint of the scoring function and improves the \ac{ahd} throughput, making scoring multiple contexts an expensive overkill.

Another dimension of the generalization versus simulation time trade-off concerns, in the LDPC case, the number of \acp{tb} to be decoded by the generated heuristics. More blocks reduce statistical noise but increase evaluation costs.
Upon experimenting, we opt for 30 \acp{tb}---fixed to ensure comparability across \ac{ahd} generations---in the target single context.

Finally, in the LDPC decoding use case, we do not find space for \textit{specialization as a feature}, as using different (meaningful) contexts to score the programs generated by our \ac{ahd} implementation yields equivalent heuristics that prove to stay valid across all other contexts.

\subsection{System-Level Evaluation Settings}

We embed and assess all \ac{cnu} functions
in a full link-level simulation based on Sionna~\cite{hoydis2023sionnaopensourcelibrarynextgeneration} so as to mirror a realistic 5G NR transport-block processing chain. 
Starting from random \ac{tb}s, the pipeline applies CRC attachment, segmentation into code blocks, \ac{ldpc} encoding with 5G NR base graphs, rate matching, modulation, transmission over a noisy channel, soft demapping, and finally \ac{ldpc} decoding with early stopping%
\footnote{Sionna does not natively include an option for early-stopping \ac{ldpc} 5G NR decoding, hence we integrated that functionality.}
governed by CRC. The candidate \ac{cnu} function is used whenever the decoder processes a parity check; all other operations, including variable-node updates and stopping rules, are preserved. This guarantees that the evaluation is genuinely system-level: a successful heuristic must decode \ac{tb}s reliably under the same conditions as a conventional decoder, and coexist with all standard physical-layer procedures.

Finally, it is worth noting that, to address the high latency of network simulations, our \ac{ahd} implementation adopts an \textit{asynchronous parallel architecture} that utilizes a distributed pool of evaluators instead of sequential loops. While \acp{llm} generate candidate programs, multiple independent workers concurrently simulate different heuristics, maximizing throughput and ensuring a scalable discovery process. 

\begin{figure*}[!ht]
    \centering
    \includegraphics[width=1\linewidth]{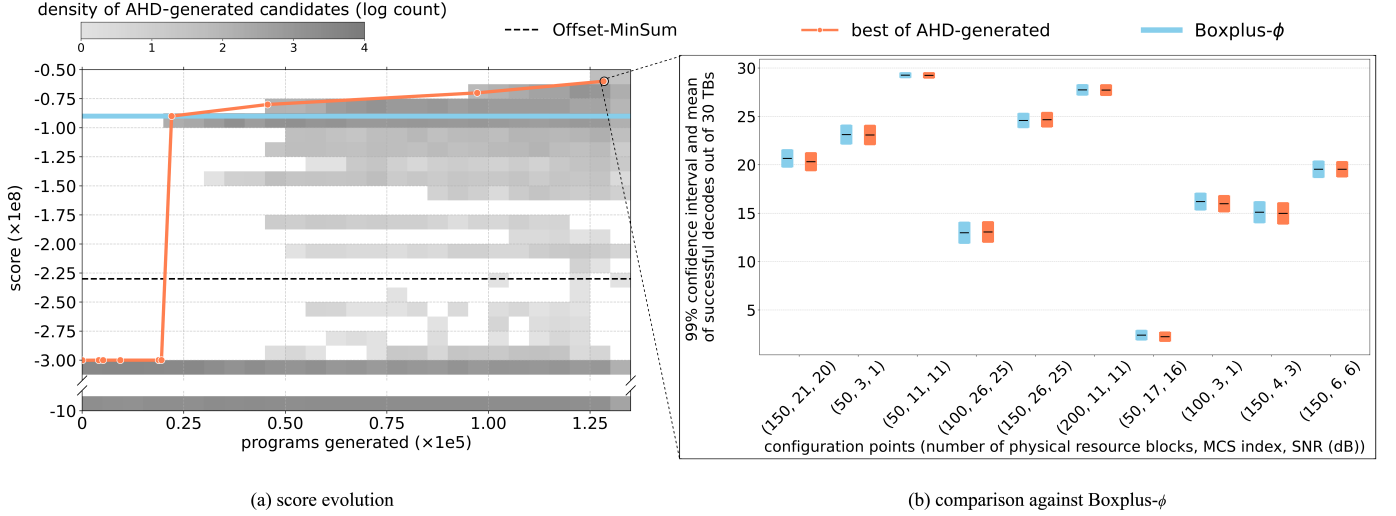}
    \caption{Results of \ac{ahd} for LDPC decoding. (a) Evolution of scores versus the number of generated programs, against the Offset Min-Sum and Boxplus-$\phi$ baselines. (b) Comparison of decoding performance, with standard deviation across \acp{tb}, between the Boxplus-$\phi$ algorithm (blue) and the \ac{cnu} function found by \ac{ahd} (orange), only for contexts where not all \acp{tb} are decoded.
    }
    \label{fig:score_ci}
    \vspace*{-8pt}
\end{figure*}

\subsection{Results and Analysis}

Fig.~\ref{fig:score_ci}(a) shows how \ac{ahd} translates the score design into decoding performance: as more programs are generated, the framework produces candidates that first gradually decode more TBs and later achieve smaller refinements in BER and needed iterations. This translates into increasing exploration of higher-score regions (gray areas) and a consequent improvement of the best score (orange). Low-score non-compilable or numerically unstable proposals still occur throughout the search as \ac{ahd} probes new directions, but are discarded by the evaluator and do not alter the overall trend. The best heuristic is found after $\approx 1.3\times 10^{5}$ generated programs; when evaluated on the 30 configuration points of Fig.~\ref{fig:heatmap}, it decodes all TBs in the contexts that are fully decodable under Boxplus-$\phi$ (purple markers in Fig.~\ref{fig:heatmap}), and in the harder contexts where not all TBs can be decoded (black markers in Fig.~\ref{fig:heatmap}) it is statistically equivalent to the state-of-the-art Boxplus-$\phi$ heuristic, as shown in Fig.~\ref{fig:score_ci}(b). In a single run, it decoded two additional TBs at one configuration point, but across 50 independent trials (each with a different set of TBs) this advantage vanishes, yielding nearly indistinguishable performance.

\begin{figure*}
    \centering
    \includegraphics[width=0.8\linewidth]{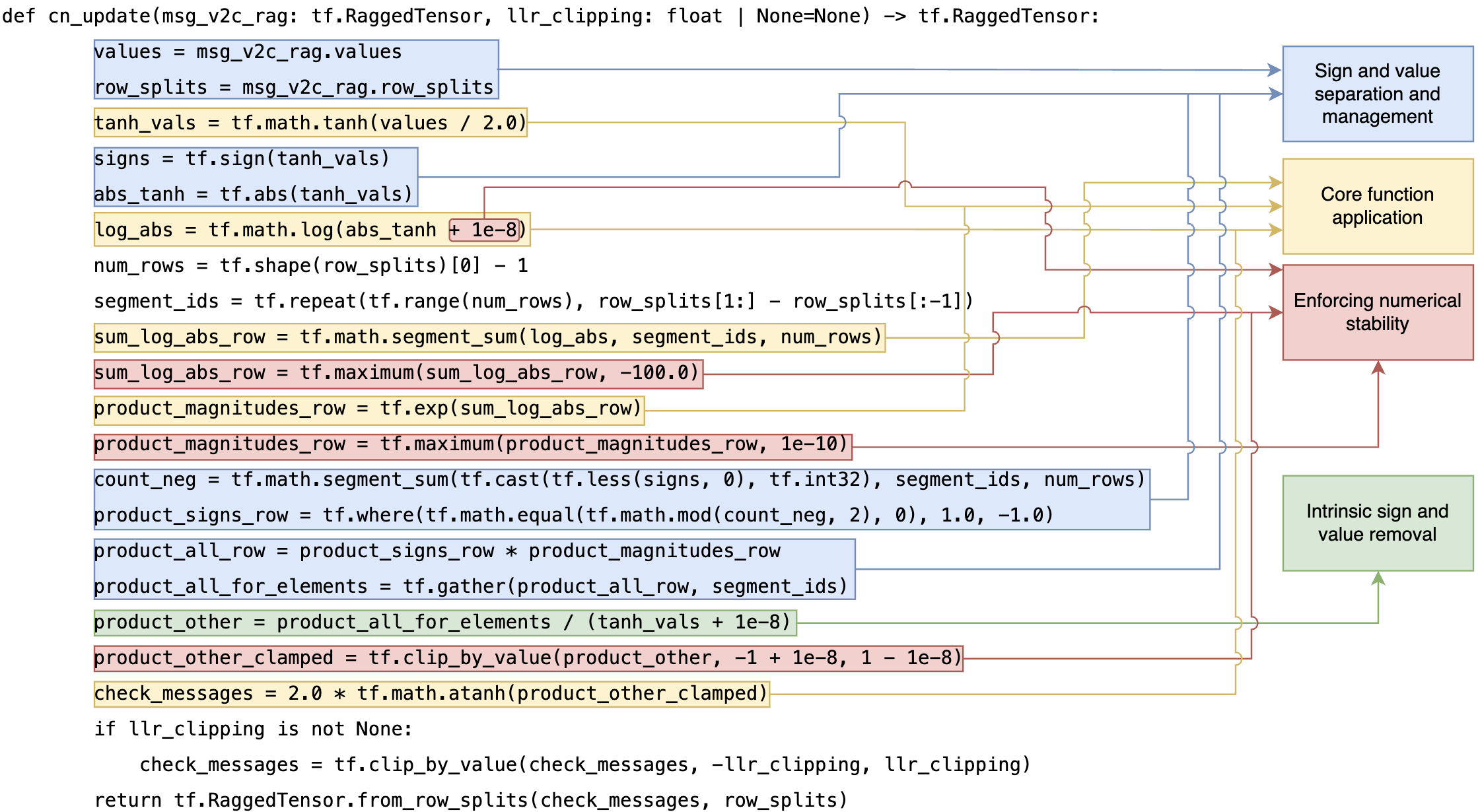}
    \caption{Best \ac{cnu} function generated by AHD, broken down into its main building blocks.}
    \label{fig:code_diff}
\end{figure*}

The actual code of the \ac{cnu} function found by AHD is shown in Fig.~\ref{fig:code_diff}. It follows the same overall procedure as the established Boxplus and Boxplus-$\phi$, but each of the four building blocks is realized differently.

In \emph{Sign and value separation and management}, AHD first maps incoming LLRs through the \texttt{tanh} nonlinearity and only then separates sign and magnitude in that \texttt{tanh} domain; Boxplus-$\phi$ instead separates sign immediately from the raw \acp{llr} before any nonlinear mapping, while the \texttt{tanh}-based Boxplus does not explicitly separate sign and magnitude because both are implicitly carried inside the signed \texttt{tanh} values and their product.

In \emph{Core function application}, AHD stays in the \texttt{tanh} and \texttt{atanh} formulation but computes the per-check-node product magnitude via a log-accumulation path (sum of log-magnitudes followed by exponentiation), whereas Boxplus forms the same aggregate through a direct product; Boxplus-$\phi$ changes domain entirely by mapping magnitudes through the $\phi$ function, aggregating by summation, and applying the $\phi$ mapping again to return to the \ac{llr} domain.

In \emph{Intrinsic sign and value removal}, AHD excludes the intrinsic edge by forming a per-node ``product of all'' quantity and then dividing out the current edge in a controlled way, while Sionna's Boxplus relies on an inversion-based workaround (multiplying by the reciprocal of each edge after computing the full product), and Boxplus-$\phi$ performs intrinsic removal as a sum-excluding-self operation in the $\phi$ domain.

Finally, in \emph{Enforcing numerical stability}, AHD concentrates its hard constraints late in the pipeline and largely at the aggregated level, stabilizing the per-node product computation and clamping only at the \texttt{atanh} input boundary; instead, Boxplus-$\phi$ enforces stability earlier by clipping inside the $\phi$ mapping, which saturates very large magnitudes before aggregation, and Boxplus introduces additional bookkeeping steps around near-zero handling to keep direct products and inversions well-behaved.

%

\section{Conclusion}

We adapt and extend LLM-based \ac{ahd} to the complex and stochastic domain of network operations. 
Our four-phase methodology comprises suitability analysis, atomic function selection, scoring design, and evaluation, and addresses the limitations of existing frameworks intended for static combinatorial problems. 
Our implementation of this framework for 5G NR LDPC decoding demonstrates that AHD can discover novel, interpretable \ac{cnu} functions with performance on par with state-of-the-art solutions implemented in production.

Ultimately, this paper makes a first step towards \ac{ahd} for network operation, with early results that show the promise of the approach. Clearly, significant future work is needed to further tailor \ac{ahd} to solve complex networking problems, so as to produce heuristics that substantially outperform current references in industry-grade target hardware platforms.
\section*{Acknowledgment}
This work was supported by the ORIGAMI project (GA 101139270) funded by SNS JU and the European Union.
L.E.~Chatzieleftheriou is funded by the European Union, 
Grant Agreement No. 101155506 (RIXISAC).
\balance

\bibliographystyle{IEEEtran}
\bibliography{references}


  \makeatletter
  \renewenvironment{IEEEbiographynophoto}[1]{%
    \normalfont\@IEEEcompsoconly{\sffamily}\footnotesize\interlinepenalty500%
    \vskip 0.5\baselineskip%
    \parskip=0pt\par%
    \noindent\textbf{#1\ }\@IEEEgobbleleadPARNLSP}{\relax\par\normalfont}
  \makeatother
\section*{Biographies}
\begin{IEEEbiographynophoto}{Reza Namvar} is a PhD student at IMDEA Networks Institute and UC3M. He holds an MSc in Software Engineering from Shiraz University and a BSc in Computer Engineering from Azad University.
\end{IEEEbiographynophoto}
\begin{IEEEbiographynophoto}{Jose Gallego} is a PhD student at IMDEA Networks Institute and UC3M. He holds a Bachelor's in Physics from UCM and a Master's in Computational Engineering and Mathematics from URV. 
\end{IEEEbiographynophoto}
\begin{IEEEbiographynophoto}{Jose A. Ayala-Romero} received his Ph.D. degree from the Technical University of Cartagena, Spain, in 2019. Currently, he is a senior researcher with the 6G Network group at NEC Laboratories Europe.
\end{IEEEbiographynophoto}
\begin{IEEEbiographynophoto}{Livia Elena Chatzieleftheriou} is an MSCA Postdoctoral fellow and a Juan de la Cierva awardee, currently working with TUDelft. She holds an M.Sc. in applied mathematics and a Ph.D. in Computer Science.
\end{IEEEbiographynophoto}
\begin{IEEEbiographynophoto}{Andres Garcia-Saavedra} is a Principal Research Scientist at NEC Laboratories Europe. He received his Ph.D. from UC3M in 2012 and works on AI-driven wireless systems, publishing and serving on TPCs for venues such as ACM MobiCom and IEEE INFOCOM, and journals including IEEE TMC and IEEE TWC.
\end{IEEEbiographynophoto}
\begin{IEEEbiographynophoto}{Albert Banchs} is the Director of IMDEA Networks Institute and Professor at Universidad Carlos III de Madrid. He earned his PhD (2002) from UPC BarcelonaTech, and previously worked at Telefónica and NEC Europe.
\end{IEEEbiographynophoto}
\begin{IEEEbiographynophoto}{Marco Fiore} is a Research Professor at IMDEA Networks Institute and co-founder at Net AI Tech Ltd. He is a Senior Member of IEEE and ACM. His research at the interface of mobile networks and data science has received recognitions that include two best paper awards at IEEE INFOCOM.
\end{IEEEbiographynophoto}

\end{document}